\documentclass[aps,superscriptaddress,unsortedaddress,prl,preprint]{revtex4-1}
\usepackage{graphicx,color} 
\begin{document}
\title{Superconducting fluctuations in organic molecular metals enhanced by Mott criticality}

\author{Moon-Sun Nam}
\affiliation{Clarendon Laboratory, Department of Physics, University of Oxford, OX1 3PU, UK}
\author{C\'ecile M\'ezi\`ere}
\author{Patrick Batail}
\affiliation{L'UNAM Universit\'e, Universit\'e d'Angers, CNRS UMR 6200, Laboratoire MOLTECH-Anjou, 2 Boulevard Lavoisier, 49045 Angers, France}
\author{Leokadiya Zorina}
\author{Sergey Simonov}
\affiliation{Institute of Solid State Physics RAS, 142432 Chernogolovka MD, Russia}
\author{Arzhang Ardavan}
\affiliation{Clarendon Laboratory, Department of Physics, University of Oxford, OX1 3PU, UK}

\begin{abstract}
Unconventional superconductivity typically occurs in materials in which a small change of a parameter such as bandwidth or doping leads to antiferromagnetic or Mott insulating phases. As such competing phases are approached, the properties of the superconductor often become increasingly exotic. For example, in organic superconductors and underdoped high-$T_\mathrm{c}$ cuprate superconductors a fluctuating superconducting state persists to temperatures significantly above $T_\mathrm{c}$. By studying alloys of quasi-two-dimensional organic molecular metals in the $\kappa$-(BEDT-TTF)$_2$X family, we reveal how the Nernst effect, a sensitive probe of superconducting phase fluctuations, evolves in the regime of extreme Mott criticality. We find strong evidence that, as the phase diagram is traversed through superconductivity towards the Mott state, the temperature scale for superconducting fluctuations increases dramatically, eventually approaching the temperature at which quasiparticles become identifiable at all.

\end{abstract}

\maketitle


Intriguing unconventional superconducting states, including those occuring in the high-$T_\mathrm{c}$ cuprates, pnictides, heavy Fermion superconductors and quasi-two-dimensional (Q2D) organic molecular superconductors, often compete with Coulomb-interaction driven antiferromagnetic or Mott insulating states~\cite{Imada-review,PAL-review}. In many cases, the parameter controlling the competition between the magnetic insulating and superconducting states is band filling; for example, high-$T_\mathrm{c}$ superconductivity occurs when the Mott insulating parent compounds such as La$_{2}$CuO$_{4}$ are doped with acceptors such as Sr on the La sites~\cite{Imada-review,PAL-review}. In the case of organic molecular superconductors, however, the first order Mott-superconductivity phase boundary is traversed by tuning the ratio, $t/U$, of the electronic bandwidth to the strength of Coulomb repulsions~\cite{Ishiguro}. In the classic case of the Q2D $\kappa$-(BEDT-TTF)$_{2}$X family of organic molecular metals, $t/U$ can be adjusted by varying the lattice spacing (and therefore the intermolecular hopping energy scale), either continuously by applying hydrostatic pressure to a Mott-insulating parent compound such as X=Cl, or sampling discretely by varying the anion volume. 
{
[In fact, in $\kappa$-(BEDT-TTF)$_{2}$X, there is a degree of frustration determined by a second hopping energy scale, $t'$, with important consequences for proximity to and stability of the Mott state~\cite{Powell,kokalj-prl-2013,kandpal-prl-2009}. However, in the following discussion, the key physics is encapsulated in the ratio $t/U$.]}
{
The} extraordinary  degree of control over the fundamental competing energy scales makes $\kappa$-(BEDT-TTF)$_{2}$X a particularly appealing experimental environment for studying strong correlations in Q2D~\cite{kanoda-review}.

Traditionally, the transition between superconducting and normal states with temperature is identified with the closure of the mean-field gap and the disappearance of the amplitude of the order parameter~\cite{Tinkham}. However, in strongly correlated superconductors (those for which Coulomb interactions influence the electronic state from which superconductivity forms), the structure of the normal state may influence the nature of this transition. Indeed, it was recognised some time ago that the observed transition temperatures for a range of unconventional superconductors is found to coincide with the temperature scale at which thermal fluctuations of the superconducting phase might be expected to destroy the long-range phase coherence~\cite{Emery,Homes,pratt-prl-2005}. This motivates a picture of an alternative kind of transition out of the superconducting state, into a state which does not exhibit the signatures of long range phase coherence such as zero resistance, but which does preserve certain features such as a gapped or pseudo-gapped spectrum, a degree of quasiparticle pairing, and remnants of the vortex physics. Experimental support for this picture was offered by measurements of magnetothermoelectric effects in under-doped high-$T_\mathrm{c}$ superconductors, which revealed a Nernst signal (a voltage transverse to both an applied temperature gradient and magnetic field) characteristic of mobile superconducting vortices persisting to temperatures well above the zero magnetic field resistive transition temperature~\cite{Ong-prb}. This was interpreted as evidence of a phase-fluctuating superconducting state.

{
However, it has also been argued that Gaussian fluctuations of the amplitude of the order parameter can give rise to an enhanced Nernst effect at temperatures above $T_\mathrm{c}$~\cite{USH,U-PRB,Serbyn}. This model offers a good quantitative account of the Nernst effect in a range of systems including amorphous superconducting thin films~\cite{Pourret,Behnia-film} and optimally- and over-doped high-$T_\mathrm{c}$ superconductors~\cite{Taillefer,Cooper}, though it is less successful for under-doped high-$T_\mathrm{c}$ superconductors, perhaps owing to the proximity of the Mott state~\cite{Taillefer}.
}

The Nernst effect 
has also been studied
in $\kappa$-(BEDT-TTF)$_{2}$X organic molecular superconductors with X=Cu(NCS)$_{2}$, for which $t/U$ is large and the normal state is Fermi-liquid-like, and with X=Cu[N(CN)$_{2}$]Br, for which $t/U$ is smaller, the metallic state is more strongly correlated, and the system is closer to the Mott transition~\cite{msn-nature}. The compound for which $t/U$ is large shows the behaviour expected for a simple superconductor to normal transition: the Nernst coefficient is large, positive and magnetic-field dependent in the vortex liquid state, but above the resistive transition temperature the Nernst coefficient is small and magnetic field independent, consistent with a normal Fermi liquid state. However the smaller-$t/U$ compound exhibits a vortex-liquid-like Nernst signal at temperatures up to 50\% above the resistive transition temperature.
{
The superconducting coherence lengths and the conductivities of these two compounds are broadly comparable, and the prediction of the Gaussian fluctuation model is that the contributions to the Nernst effect from superconducting fluctuations should also be comparable~\cite{USH,U-PRB}. The significant difference in Nernst effects between the smaller-$t/U$ and the larger-$t/U$ compounds was taken as evidence for a phase-fluctuating superconducting state occurring in the proximity of the Mott state~\cite{msn-nature}.
}

\begin{figure}[!htp]
\includegraphics[width=12cm]{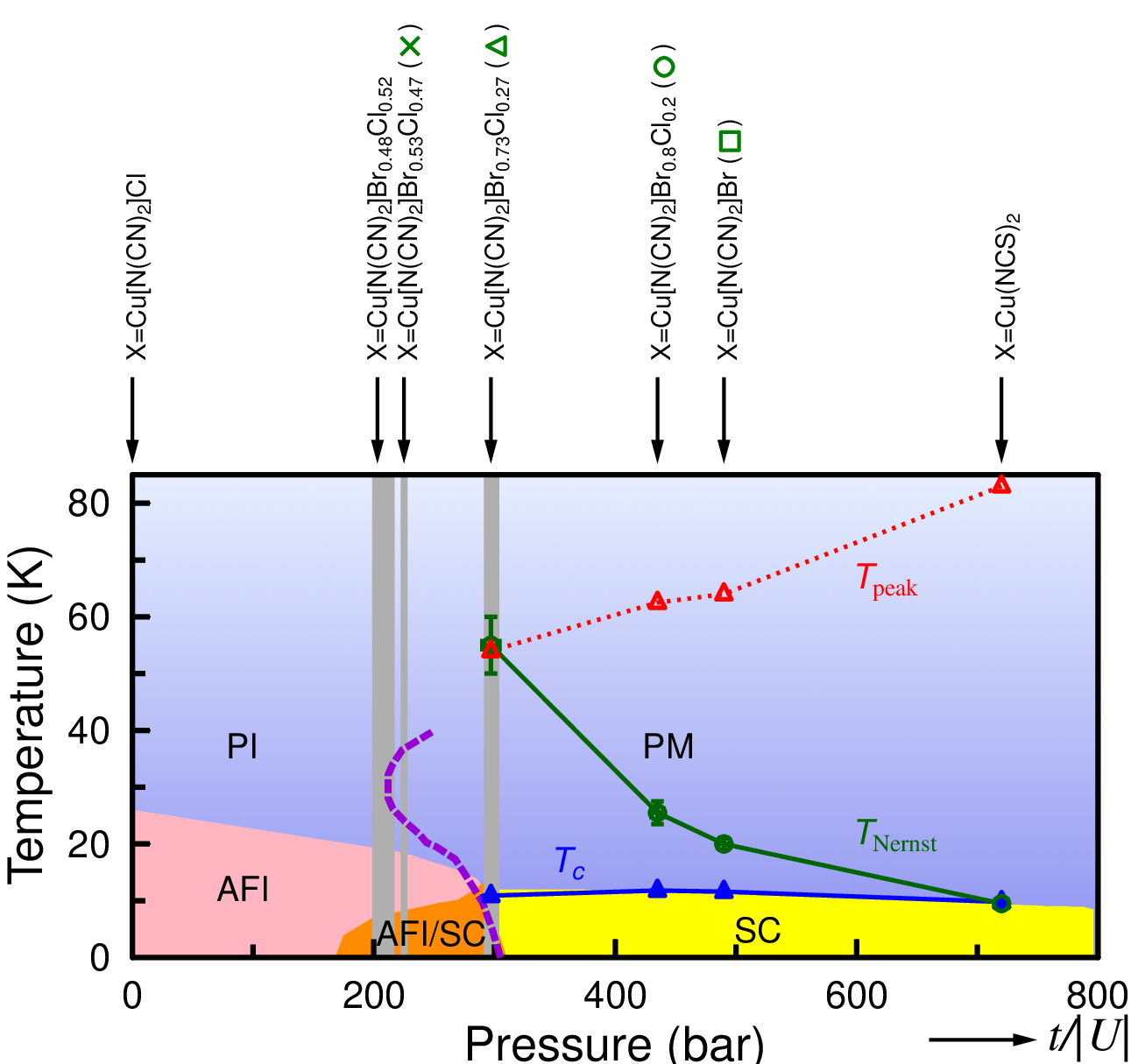}
\caption{\textbf{Phase diagram of $\kappa$-(BEDT-TTF)$_{2}$X.} The phase diagram as a function of temperature and pressure, or, equivalently, $t/U$.  The location of each compound $\kappa$-(BEDT-TTF)$_{2}$X with different anion, X, is indicated on the effective pressure scale by a grey line whose width indicates the degree of uncertainty in effective pressure, based on electrical transport and superconducting properties. X=Cu(NCS)$_{2}$ data are taken from Ref.~\cite{msn-nature}. The dashed Mott critical line is taken from Ref.~\cite{NMR-Kagawa}.
PI: paramagnetic insulator; PM: paramagnetic metal; AFI: antiferromagnetic insulator; SC: superconductor }
\label{phase-dia}
\end{figure}

Together, the {
Nernst} observations in two very different classes of strongly correlated superconductors, 
{
namely, under-doped high-$T_\mathrm{c}$ superconductors and $\kappa$-(BEDT-TTF)$_{2}$X organic molecular superconductors} 
support the conjecture that superconducting states lying close to a Mott transition are particularly susceptible to fluctuations. In this work, we seek to explore how the superconducting fluctuations develop in $\kappa$-(BEDT-TTF)$_{2}$X as the Mott state is approached by reducing $t/U$.  
Perhaps the most elegant way of addressing this question experimentally is to begin with a compound, such as X=Cu[N(CN)$_{2}$]Cl, that lies within the Mott state at ambient pressure (see Figure~\ref{phase-dia}). Applying hydrostatic pressure tunes $t/U$ continuously as the lattice parameters shrink: at low pressures the antiferromagnetic ordering temperature is suppressed somewhat; at about 300 bar, there is evidence from NMR~\cite{NMR-Lefebvre,NMR-Kagawa}, resistance measurements~\cite{resistance-Limelette,critical-Kagawa,transport1-Kagawa,transport2-Kagawa}, and ultrasonic velocity measurement~\cite{ultrasonic} for a mixed Mott insulating and superconducting phase; above about 300 bar, the ground state is superconducting; and as the pressure increases further, the transition temperature falls as the band width increases and the normal state density of states decreases~\cite{Caulfield}. 

With temperature, these measurements reveal a first order transition (purple dashed line in figure ~\ref{phase-dia}), with a critical end point at a temperature of around 40~K and a pressure of about 250~bar. A very detailed scaling analysis of the conductance in the vicinity of the critical end point revealed that the critical exponents for this model Q2D Mott transition do not correspond to those for any previously-identified universality class~\cite{critical-Kagawa}, motivating further the study of the properties of the adjoining unconventional superconducting state.

While it is possible to study thermoelectric phenomena under pressure~\cite{weida-pf6-nernst}, the presence of a pressure-transmitting medium leads to uncertainties about heat transport paths that are not present for experimental configurations in which the sample is in a vacuum. For this reason, we chose to tune $t/U$ instead by alloying the compounds X=Cu[N(CN)$_{2}$]Cl and X=Cu[N(CN)$_{2}$]Br. These two anions are isoelectronic but the latter is slightly smaller in volume, so the electronic configuration is independent of the composition but the lattice parameters, and hence $t/U$, do depend on composition. In principle, by selecting the alloy composition, it is possible to synthesise a crystal exhibiting any value of $t/U$ within the range determined by those for the two pure compounds.

\section{Results}

\begin{figure}[!htp]
\includegraphics[width=10cm]{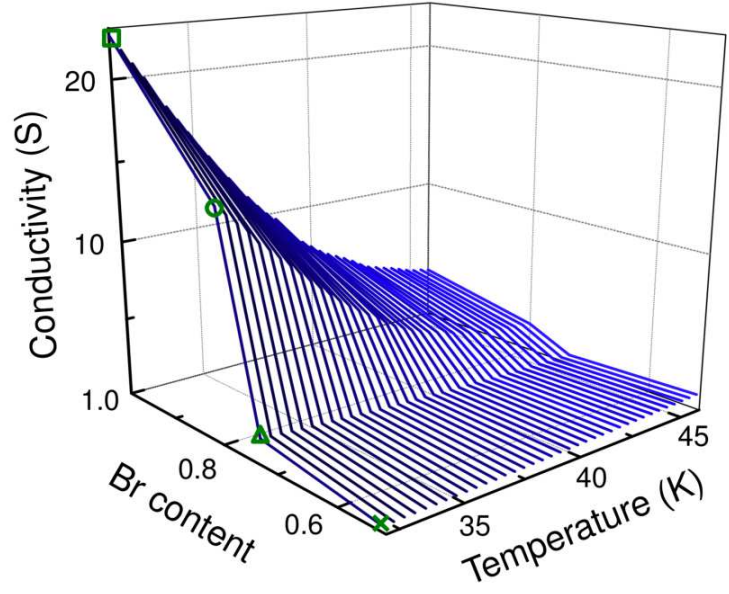}
\caption{\textbf{Conductivity as a function of temperature and Br content in $\kappa$-(BEDT-TTF)$_{2}$Cu[N(CN)$_{2}$]Cl$_{1-x}$Br$_{x}$.} The conductivity as a function of temperature and effective pressure is shown, sampled at four effective pressures: open square for Br$_{1.0}$, open circle for Br$_{0.8}$, open triangle for Br$_{0.73}$, and cross for Br$_{0.53}$. The very abrupt decrease in conductivity between Br$_{0.8}$ and Br$_{0.73}$ is an indication of approaching Mott criticality as seen in other compounds such as $\kappa$-(BEDT-TTF)$_{2}$Cu[N(CN)$_{2}$]Cl~\cite{critical-Kagawa}  and V$_{2}$O$_{3}$~\cite{Limelette-science}. }
\label{conductivity} 
\end{figure}

Table~1 lists the compositions of the four alloys, X=Cu[N(CN)$_{2}$]Cl$_{1-x}$Br$_{x}$, and the pure X=Cu[N(CN)$_{2}$]Br compound that we studied. Their conductivities as a function of temperature are plotted in Figure~\ref{conductivity} in a form that facilitates comparison with Figure~1c of Reference~\cite{critical-Kagawa}, but replacing the pressure axis with composition; this plot demonstrates that our compositions span the Mott critical region, with the Br$_{1.0}$ and Br$_{0.8}$ alloys becoming highly conductive at low temperatures, but the Br$_{0.73}$ and Br$_{0.53}$ remaining much more resistive.

\begin{figure}[!htp]
\includegraphics[width=10cm]{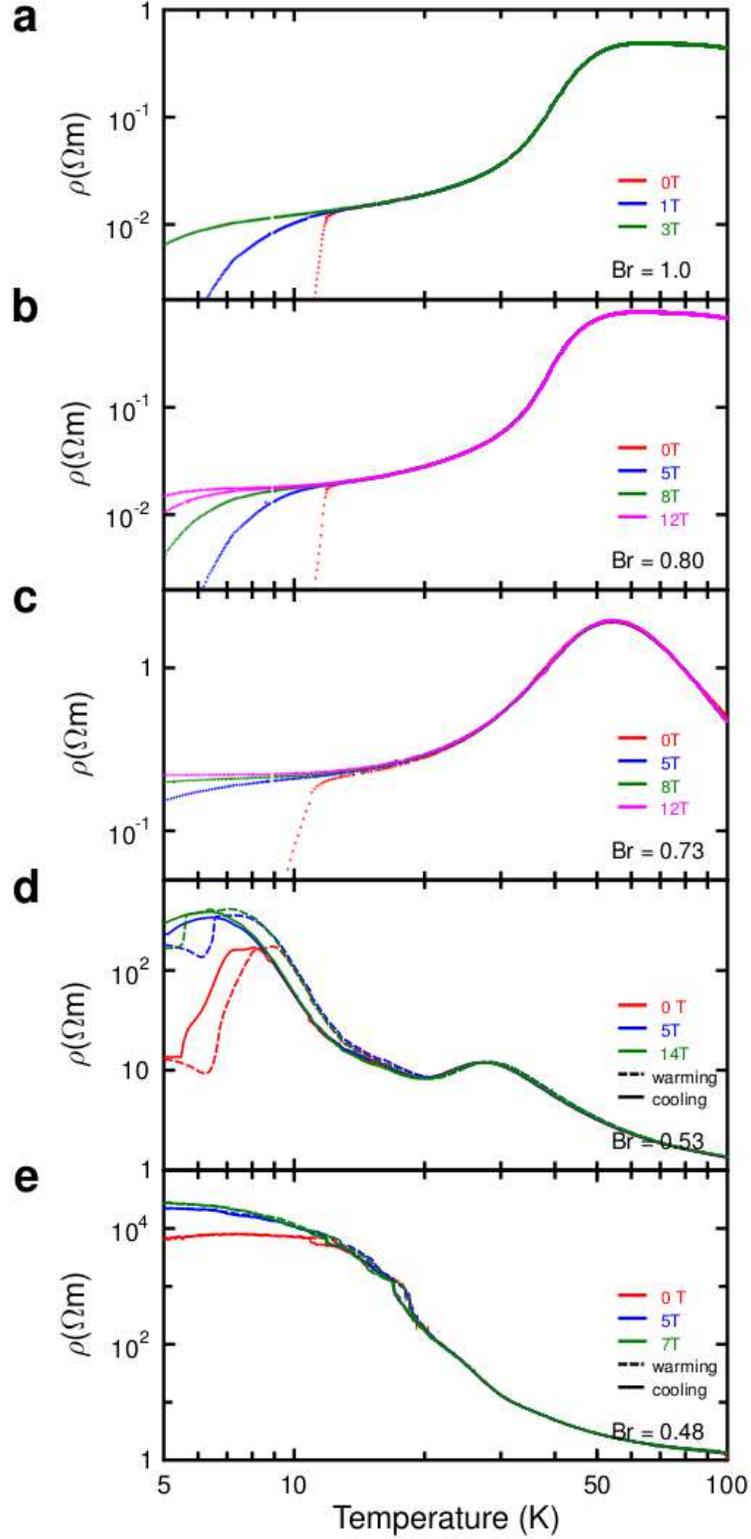}
\caption{\textbf{Logarithmic plots of resistance as a function of temperature in $\kappa$-(BEDT-TTF)$_{2}$Cu[N(CN)$_{2}$]Cl$_{1-x}$Br$_{x}$.} \textbf{a,} Br$_{1.0}$ at 0, 1, and 3~T. \textbf{b,} Br$_{0.8}$ at 0, 5, 8 and 12~T. \textbf{c,} Br$_{0.73}$ at 0, 5, 8 and 12~T. \textbf{d,} Br$_{0.53}$ at 0, 5, and 14~T. \textbf{e,} Br$_{0.48}$ at 0, 5, and 7~T.}
\label{log-log-RvsT}
\end{figure}

Log-log plots of {
interlayer} resistivity as a function of temperature with different magnetic fields applied perpendicular to the Q2D layers are shown in Fig~\ref{log-log-RvsT}. They reveal the range of different behaviours expected as the Mott transition is traversed, and allow us to estimate the calibration between alloy composition and effective pressure. 

The superconducting Br$_{1.0}$ (with the largest $t/U$ in this study), Br$_{0.8}$, and Br$_{0.73}$ compounds each show metallic conductivity below 64~K, 62~K and 54~K respectively. It is a general characteristic of superconducting members of the $\kappa$-(BEDT-TTF)$_{2}$X family that they exhibit a crossover in this temperature range from a regime of diffusive conductivity characterised by negative ${\mathrm d}R/{\mathrm d}T$ at high temperatures to a regime in which the electronic transport is coherent and metallic (with positive ${\mathrm d}R/{\mathrm d}T$) at lower temperatures. This crossover is well described by dynamical mean field theory~\cite{resistance-Limelette,DMFT-Mckenzie,deng-kotliar-prl, Ben-coh} and 
{
$T_\mathrm{peak}$}, 
the temperature at which the peak in the resistivity occurs, increases with $t/U$. (It also exhibits a dependence on certain classes of disorder~\cite{disorder-cuscn}.) 
{
Note that the metallic compounds that we study here do not all exhibit simple $T^2$-dependence of resistivity over significant temperature ranges~\cite{yasin-epl}, so we adopt the temperature at which the resistance peaks as a consistent phenomenological measure of the temperature below which coherent transport occurs.}

The temperature dependence of the resistivity in Br$_{0.8}$ is rather similar to that of Br$_{1.0}$, but the coherent transport crossover occurs at a slightly lower temperature, the superconducting transition temperature is slightly higher, and the resistivity is almost a factor of two higher throughout. This places Br$_{0.8}$ close to, but a little lower than, Br$_{1.0}$ on the $t/U$ scale.

Optical studies~\cite{Dressel-prb1,Dressel-prb2} indicate that the compound with X=Cu[N(CN)$_{2}$]Cl$_{0.3}$Br$_{0.7}$ lies at the boundary of the Mott state at low temperatures. 
Consistent with this, the absence of hysteresis in the resistivity of Br$_{0.73}$ indicates that it lies above the Mott transition in $t/U$. However, compared with Br$_{1.0}$ and Br$_{0.8}$, the coherent transport crossover temperature $T_\mathrm{peak}$ is significantly suppressed in Br$_{0.73}$, and its normal state resistivity is over an order of magnitude higher, placing Br$_{0.73}$ substantially below Br$_{1.0}$ and Br$_{0.8}$ on the $t/U$ scale and very close to the Mott transition.
{
Optical studies~\cite{Dressel-prb1,Dressel-prb2} reveal that Br$_{0.73}$ exhibits a substantially enhanced effective mass and scattering rate compared to compounds with higher Br content.}

By contrast with the superconducting compounds, Br$_{0.53}$ shows insulating behaviour over most of the temperature range. It shows a small region of positive ${\mathrm d}R/{\mathrm d}T$  between about 20~K and 30~K which we believe is a vestige of the crossover towards coherent electronic transport, but below 20~K the negative ${\mathrm d}R/{\mathrm d}T$ is recovered. Following the analysis in References~\cite{resistance-Limelette} and ~\cite{transport1-Kagawa}, we find that the resistivity follows approximately activated behaviour between about 30~K and 50~K, yielding a charge gap of $\Delta \sim 100 \pm 1.5$~K. This is comparable with the behaviour of the pure X=Cu[N(CN)$_{2}$]Cl compound under a hydrostatic pressure of $225\pm3$~bar~\cite{resistance-Limelette,transport1-Kagawa}. Below about 7.5~K, Br$_{0.53}$ exhibits a dramatic resistivity down-turn. The resistivity is highly hysteretic with temperature in this region, consistent with the ground state of Br$_{0.53}$ lying in the mixed phase, and the resistivity down-turn arising from the occurrence of regions of superconductivity.

The Br$_{0.48}$ sample (with the smallest $t/U$ in this study) is highly resistive and shows insulating behaviour (${\mathrm d}R/{\mathrm d}T$ is negative) over the whole temperature range except below about 7~K where there is a down-turn in the resistivity, though this is much weaker than in Br$_{0.53}$. There is little magnetic field dependence, other than to suppress the weak resistivity down-turn at low temperatures. Br$_{0.48}$ has a charge gap of $130\pm20$~K. This is comparable with the behaviour of the pure X=Cu[N(CN)$_{2}$]Cl compound under a hydrostatic pressure of $210\pm10$~bar; the magnetic-field-dependent weak resistivity down-turn suggests that, like Br$_{0.53}$, it lies in the mixed-phase region, but that the superconducting fraction is much smaller. 

We note that the locations of the alloys on the phase diagram that we derive from this analysis are consistent with what is found in other experimental studies of these alloys~\cite{Dressel-prb1,Dressel-prb2, yasin-epl}.

\begin{figure}[!htp]
\includegraphics[width=14cm]{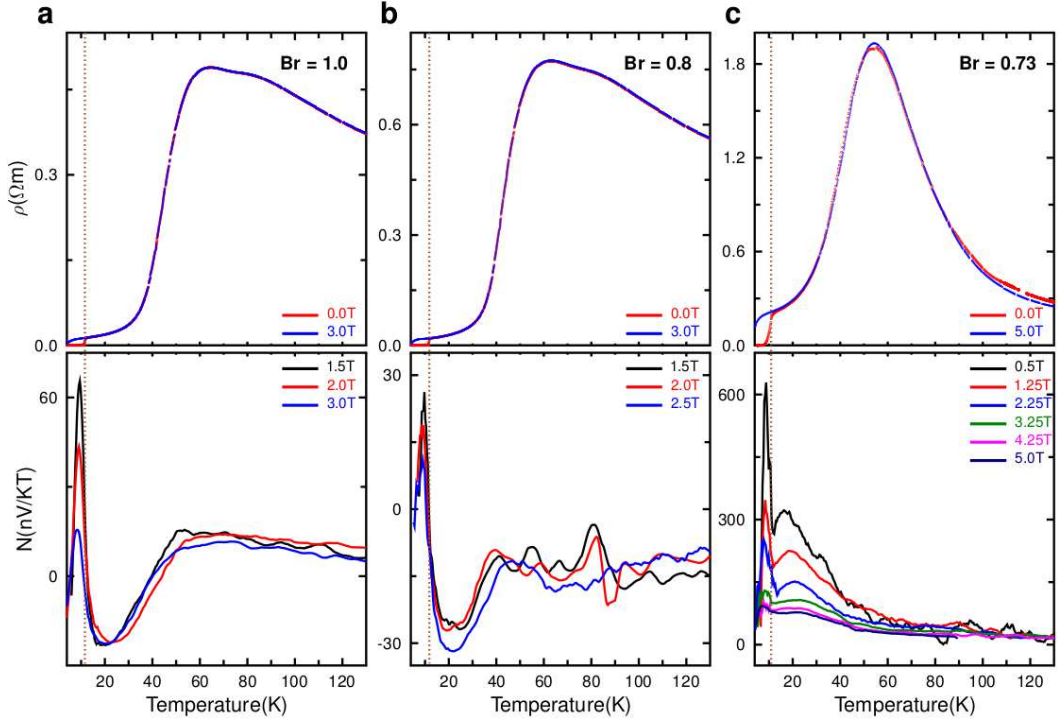}
\caption{\textbf{The Nernst effect in $\kappa$-(BEDT-TTF)$_{2}$Cu[N(CN)$_{2}$]Cl$_{1-x}$Br$_{x}$.}  Resistivity as a function of temperature at various fixed magnetic fields  are also shown for each alloy. The dashed line indicates the superconducting transition at 0T. \textbf{a,} Br$_{1.0}$ resistivity at 0 and 3~T and Nernst at 1.5, 2.0 and 3.0~T.  \textbf{b,} Br$_{0.8}$ resistivity at 0 and 3~T and Nernst at 1.5, 2.0 and 2.5~T. \textbf{c,} Br$_{0.73}$  resistivity at 0 and 5~T and Nernst at 0.5, 1.25, 2.25, 3.25, 4.25 and 5.0~T. }
\label{Nernst-resistivity}
\end{figure}

Equipped with a selection of alloys at well-defined positions on the $t/U$ axis, we are now in a position to investigate the central theme of this work: how the superconducting fluctuations, probed through the Nernst effect, depend on $t/U$. Figure~\ref{Nernst-resistivity} shows the resistivity and the Nernst coefficient, $N$, (defined by $E_y = -NB_z\nabla_xT$, where $E_y$ is the measured electric field in response to applied temperature gradient $\nabla_xT$ and magnetic field $B_z$) as a function of temperature for a range of fixed magnetic fields for each of the materials that exhibits metallic behaviour. 
{
The resistivity was measured in the interlayer direction (along the $c$-axis); the temperature gradient was applied in the intralayer direction (parallel to the $a$--$b$ plane); the Nernst signal was measured in the intralayer direction perpendicular to the temperature gradient (see Methods).}

In each case at high temperatures, in the regime in which the resistivity indicates that the electrical transport is diffusive~\cite{DMFT-Mckenzie,deng-kotliar-prl}, the Nernst coefficient is small and essentially temperature independent.
As the temperature falls, and coherent transport of quasiparticles is established, the character of the Nernst signal changes. For Br$_{1.0}$ and Br$_{0.8}$, the coherent quasiparticle signal becomes larger in magnitude and negative. As the resistive superconducting transition is approached the Nernst signal acquires a large, positive, and magnetic-field-dependent component which peaks within the superconducting state and then falls as the temperature falls further. The Nernst effects in Br$_{1.0}$ and Br$_{0.8}$ are very similar to each other, and both are consistent with the results reported in Ref.~\cite{msn-nature} in which the behaviour of Br$_{1.0}$ in the vicinity of the superconducting transition was examined in detail.

The Br$_{0.73}$ alloy shows qualitatively different behaviour from Br$_{1.0}$ and Br$_{0.8}$ at temperatures below which coherent quasiparticle transport is established: it exhibits a large, positive and magnetic-field dependent signal for all temperatures below the resistivity downturn. The resistivity does not show any magnetic field dependence above $T_\mathrm{c}$, implying that the magnetic field dependence of the Nernst signal has its origin in the thermoelectric tensor.

\section{Discussion}

In a single-parabolic-band isotropic metal, the Nernst coefficient is expected to vanish~\cite{sondheimer}, but anisotropic materials or those in which multiple bands cross the Fermi energy may exhibit significant Nernst effects, particularly if the quasiparticle mobility is high~\cite{Behnia-JPCM}. In weak magnetic fields (i.e.\ when the magnetic field energy scale is small compared to the Fermi energy) the Nernst signal generated by a Fermi liquid is linear in magnetic field and the Nernst coefficient, $N$, is field-independent.

In contrast, a large and magnetic-field-dependent Nernst coefficient is expected in the vortex liquid state of a superconductor. The normal cores of the vortices, being more efficient carriers of heat than the surrounding gapped superfluid, flow down a temperature gradient applied perpendicular to the magnetic field. In doing so, they wind a superfluid phase in the direction perpendicular to both the magnetic field and the temperature gradient, at a rate that is proportional to their density (i.e.\ the magnetic field strength) and to their speed (which is determined by the temperature gradient and the vortex mobility). If the average separation of vortices is comparable to or smaller than the penetration depth, the vortices interact repulsively, impeding their mobility. Thus, in $\kappa$-(BEDT-TTF)$_{2}$X, the vortex Nernst coefficient is expected to be large but to decrease with magnetic field for fields above a small fraction of 1~T~\cite{msn-nature}. 


The onset of a positive and magnetic-field-dependent Nernst coefficient at temperatures below about 18~K (1.5 times $T_\mathrm{c}$) in Br$_{1.0}$ has previously been identified as evidence of a phase-fluctuating superconducting state~\cite{msn-nature}, that, while unable to exhibit a zero-resistance state owing to the loss of long-range phase coherence, is nevertheless able to support vortex-like excitations owing to sufficiently strong shorter-range phase correlations. The data in Fig.~\ref{Nernst-resistivity} indicate that, although it lies closer to the Mott state on the $t/U$ scale than Br$_{1.0}$, Br$_{0.8}$ behaves in a very similar way, with the onset of a vortex-like Nernst signal appearing instead below about 20~K (1.8 times $T_\mathrm{c}$).

Br$_{0.73}$, judging by the temperature below which it exhibits coherent quasiparticle transport and its overall higher resistivity, lies much closer to the Mott transition on the $t/U$ scale than Br$_{1.0}$ or Br$_{0.8}$, and its Nernst behaviour is qualitatively different. As soon as the temperature falls enough for quasiparticles characterised by coherent transport to emerge (below about 54~K) there is the onset of a very large 
{
positive} magnetic-field-dependent Nernst coefficient that evolves into the familiar vortex Nernst signal below $T_\mathrm{c}$ of 10.9~K. 
{
The overall magnitude of the Nernst signal in Br$_{0.73}$, which is expected to be proportional to the sample resistivity~\cite{Behnia-JPCM}, is larger than Nernst signals in Br$_{1.0}$ and Br$_{0.8}$, but this does not account for the decrease in the Nernst coefficient with magnetic field or for its positive sign.}

{
One} natural explanation for this observation is that the onset of superconducting 
{
phase} fluctuations occurs at higher temperatures as the Mott state is approached along the $t/U$ scale. This extends the trend observed earlier and reported in Ref.~\cite{msn-nature}. The material with X=Cu(NCS)$_{2}$, which lies substantially higher up the $t/U$ scale than any of the materials studied here, shows no signs of superconducting fluctuations above $T_\mathrm{c}$ in its Nernst effect~\cite{msn-nature}. Br$_{1.0}$ and Br$_{0.8}$, lying at successively lower points on the $t/U$ scale, exhibit the onset of a vortex-like Nernst signal below about 1.5$T_\mathrm{c}$ and 1.8$T_\mathrm{c}$ respectively. In Br$_{0.73}$, the compound exhibiting the smallest $t/U$ of the superconductors in this study, the positive and strongly field-dependent Nernst signal occurs below about 5$T_\mathrm{c}$.

{
Another possibility is that Gaussian amplitude fluctuations of the superconducting order parameter give rise to the enhanced Nernst signal above $T_\mathrm{c}$. The prediction of the standard Gaussian fluctuation theory~\cite{USH,U-PRB} is that the Nernst coefficient becomes magnetic-field-independent at low fields, and this is observed in both amorphous superconducting films~\cite{Pourret, Behnia-film} and over- and optimally-doped high-$T_\mathrm{c}$ superconductors~\cite{Taillefer, Cooper}. Our observation of a field-dependent Nernst effect down to the lowest magnetic fields rules out the possibility of a quantitative comparison of our data to the Gaussian fluctuation theory. Furthermore, the assumption that the normal state of the Br$_{0.73}$ compound is a Fermi liquid (an assumption on which this theory is based) may not be a safe one, given its proximity to the Mott transition: it does not exhibit a $T^2$ resistivity; other studies of $\kappa$-(BEDT-TTF)$_{2}$X close to the Mott transition reveal non-Fermi-liquid behaviour~\cite{NMR-Kagawa, critical-Kagawa}; and previous studies of this family of alloys show that the strong correlations in the Br$_{0.73}$ significantly modify the effective mass and the optical properties~\cite{Dressel-prb1,Dressel-prb2}. 
}

We note that, 
{
interpreting the enhanced Nernst signal as arising from superconducting fluctuations of some kind,} 
there is an interesting convergence of temperature scales. In Br$_{0.73}$, the underlying temperature scale 
below which superconducting fluctuations are observed becomes coincident with the temperature scale for coherent quasiparticle transport. This implies that as soon as the temperature becomes low enough for the excitations of the system to be identified with states of well-defined momentum, they exhibit the tendency towards pairing and phase coherence. Both temperatures are quite comparable with the temperature of the critical endpoint of the first-order Mott transition line. 

While it is true that the only mechanisms that are known to give rise to magnetic-field-dependent Nernst coefficients at low magnetic fields are related to superconductivity, this does not preclude the existence of other, as yet unidentified, mechanisms~\cite{Behnia-JPCM}. We cannot rule out the possibility that such a mechanism lies behind the Nernst behaviour in Br$_{0.73}$. Given the location of Br$_{0.73}$ on the $t/U$ scale, it seems likely that the underlying physics driving any effect of this kind would depend on the proximity of the Mott state.

\section{Methods}

Single crystals of solid solution $\kappa$-(BEDT-TTF)$_2$Cu[N(CN)$_2$]Br$_x$Cl$_{1-x}$ ($x$ = 0.48, 0.53, 0.73, 0.80, and 1) were synthesized by electrocrystallization. For the bromine compound ($x$ = 1), PPh$_4$[N(CN)$_2$] and CuBr were used as electrolyte in a 1,1,2-trichloroethane/ethanol 10\% vol. mixture; for the alloys, the electrocrystallization was carried out in the same solvent mixture containing NaN(CN)$_2$, 18-crown 6, CuBr and CuCl. The $x$ values were determined by microprobe analysis and full single crystal X-ray structure refinements. 
{
The single crystals were platelets with faces of the order of 1~mm across and thicknesses of the order of 100~$\mu$m.
}

The terminal ethylene groups of the BEDT-TTF molecule in $\kappa$-BEDT-TTF$_2$\-Cu[N\-(CN)$_{2}$]Br undergo an ordering transition, with an associated small decrease in resistivity, at about 80~K. Alloys showing metallic behaviour with coherent peaks exhibited comparable features in resistivity at around 80~K. Although in insulating alloys the effect was not as prominent as in metallic ones, in resistivity measurements all samples were cooled down at the same low rate of 150~mK$\cdot$min$^{-1}$ from 160~K to 4~K to ensure that the terminal ethylene disorder was controlled.

We made interlayer resistivity measurements using standard four-probe ac lockin techniques with a maximum current of 10~$\mu$A at a frequency of 13.7~Hz for Br$_{1.0}$, Br$_{0.8}$, and Br$_{0.73}$. For insulating Br$_{0.53}$, we used slow lockin techniques with a maximum current of 500~nA at a frequency of 0.137~Hz and a time constant 100 seconds. For insulating Br$_{0.48}$, we used a d.c.\ current of 100~nA. We repeated each temperature sweep for positive and negative d.c.\ current and took the average to remove any the thermoelectric effects.

We applied a temperature gradient by thermally coupling each end of the sample to a single-crystal quartz block using graphite paint and 50~$\mu$m gold wire. The quartz blocks were weakly thermally coupled to a copper base plate and base temperature was monitored using a cernox sensor.  The temperature gradient was controlled using 100~$\Omega$ chip resistor-heaters and measured using thermocouples.  Below 10K, the thermal gradient measured by thermocouples was checked using calibrated RuO thermometers mounted on each quartz block; there was always good agreement. We applied temperature differences of typically 500~mK in the high temperature range and 250~mK at temperatures below 40~K. The experiment was maintained under a vacuum of below $10^{-5}$~mbar to ensure thermal isolation. The magnetic field was applied perpendicular to the BEDT-TTF conducting planes using a superconducting magnet in persistent mode to minimise electrical noise. Transverse thermal voltage contacts were made using resistive phosphor bronze wire of diameter 70~$\mu$m to reduce heat loss through these contacts. The signal was measured via oxygen-free copper wire with mechanical joints only from the low-temperature stage all the way to the Keithley voltmeter preamplifiers to minimise thermal voltage noise. The data were recorded over the range $\sim$4~K to $\sim$150~K in temperature sweeps at a rate typically of 100~mK/min.

Each temperature sweep was repeated for positive and negative magnetic fields. Longitudinal thermoelectric contributions to the voltage measured were removed by subtracting the positive and negative field sweeps. 
{
Uncertainties in the intraplane temperature profile arising from the irregularity of the sample shapes meant that it was not possible to separate intraplane components of the Nernst signal. At 1~T, the typical maximum longitudinal contribution to the signal before this correction was performed was about ten to twenty times the maximum transverse contribution. The longitudinal component was almost completely magnetic field independent above $T_{\mathrm c}$. 
}

\section{Acknowledgements}

Work at Oxford was supported by EPSRC grant EP/G003610/1. Work at Angers was supported by the joint CNRS-Russian Federation grants PICS 6028 and
RFBR-CNRS 12-03-91059.

\section{Author contributions}

M.-S.N.\ and A.A.\ formulated the investigation. C.M.\ and P.B.\ prepared and characterised the samples. L.Z.\ and S.S.\ performed X-ray structure refinements. M.-S.N.\ performed the experiments. M.-S.N.\ and A.A.\ wrote the manuscript.

\section{Competing financial interests}

The authors have no competing financial interests.

\newpage

\begin{table}[!htp]
\caption{}
\begin{center}
\begin{tabular}{|c|c||c|c|c|c|c|} \hline
Br content, $x$ & Label & $T_\mathrm{c}$ (K)&$T_\mathrm{peak} $ (K) & $T_\mathrm{Nernst}$ (K) & $\Delta$(K)  & Effective pressure (bar)  \\ \hline \hline
 1.0  & Br$_{1.0}$   & 11.6      & 64                       & $18\pm1$ &                 & 490 $\pm$ 1 \\
0.8  & Br$_{0.8}$        & 11.8       & 62                       & $20\pm2$ &                 & 435 $\pm$ 1 \\
0.73  & Br$_{0.73}$      &     10.9       & 54                       & $55\pm5$ &                 & 297 $\pm$ 4 \\ \hline
0.53  & Br$_{0.53}$      & 7.5  $\pm$ 0.5  
\footnote{The superconducting transition is not complete in these compounds; this is the temperature below which there is positive magnetoresistance at low fields.}
&          &        & 100 $\pm$ 1.5 & 225 $\pm$ 3 \\
 0.48  & Br$_{0.48}$      &     7.0$\pm$ 0.5$^{\:\mathrm a}$           &           &                   & 130 $\pm$ 20 & 210 $\pm$ 10 \\ \hline

\end{tabular}
\end{center}
\label{default}
\end{table}

\end{document}